# Wavelength-scale ptychographic coherent diffractive imaging using a high-order harmonic source

Getnet K. Tadesse[1,2,*], Wilhelm Eschen[1,2], Robert Klas[1,2], Maxim Tschernajew[1,2], Frederik Tuitje[1,3], Michael Steinert[2], Matthias Zilk[2], Vittoria Schuster[2], Thomas Pertsch[2,4], Christian Spielmann [1,3], Jens Limpert[1,2,4], Jan Rothhardt[1,2]

[1]Helmholtz-Institute Jena, Fröbelstieg 3, 07743 Jena, Germany
[2]Institute of Applied Physics, Abbe Center of Photonics, Friedrich-Schiller-University Jena, Albert-Einstein-Straße 15, 07745 Jena, Germany
[3]Institute of Optics and Quantum Electronics, Abbe Center of Photonics, Friedrich-Schiller-University Jena, Max-Wien-Platz 1, 07743 Jena, Germany
[4]Fraunhofer Institute for Applied Optics and Precision Engineering, Albert-Einstein-Str. 7, 07745 Jena, Germany
* Corresponding author, getnet.tadesse@uni-jena.de

**Ptychography enables coherent diffractive imaging (CDI) of extended samples by raster scanning across the illuminating XUV/X-ray beam thereby generalizing the unique advantages of CDI techniques. Table-top realizations of this method are urgently needed for many applications in sciences and industry. Previously, it was only possible to image features much larger than the illuminating wavelength with table-top ptychography although knife-edge tests suggested sub-wavelength resolution. However, most real-world imaging applications require resolving of the smallest and closely-spaced features of a sample in an extended field of view. In this work, we resolve features as small as 2.5 $\lambda$ (45 nm) by using a table-top ptychography setup and a high-order harmonic XUV source. For the first time, a Rayleigh-type criterion is used as a direct and unambiguous resolution metric for high-resolution table-top setup. This reliably qualifies this imaging system for real-world applications e.g. in biological sciences, material sciences, imaging integrated circuits and semiconductor mask inspection.**

Coherent diffractive imaging and related techniques are variants of X-ray microscopy that have become increasingly important for photon energies where focusing optics are limited in terms of quality and achievable spot size[1,2]. Ptychography is a natural extension of the basic CDI technique where a sample is raster scanned relative to the illuminating beam and multiple diffraction patterns are recorded[3,4]. This technique allows for imaging of extended samples and opens the door for wider application of diffractive imaging techniques[5,6]. Ptychography can also be integrated into 3D tomography to record different projections of a specimen thus enabling a complete 3D map of a sample both in amplitude and phase[7,8]. Thus, complete nanoscale 3D images of integrated circuits[8] and whole unstained cells[9] have become possible and attracted huge attention in science and technology, particularly bridging the gap between visible light and electron microscopy. Until recently, these ptychography experiments have been conducted at synchrotron facilities, which are not easily accessible. The availability of XUV and soft X-ray sources based on high-order harmonic generation (HHG)[10] and gas-discharge plasma[11] in recent years made highly accessible table-top setups possible. HHG-based table-top sources have already demonstrated a photon flux up to $10^{14}$ photons/s at 21 eV [12] and photon energies extending up to the water window and beyond[10,13,14].

Ptychography experiments using table-top sources have demonstrated resolutions of 40 nm[15] and 58 nm[16] for non-periodic samples and sub-20 nm resolutions for periodic samples[17], respectively, by applying a 10%-90% knife-edge criterion. The reported resolution values in these table-top experiments were on the order of the illuminating wavelength. However, the actual feature sizes of the samples that have been resolved were much larger (> 10$\lambda$). Most applications of high-resolution imaging, however, require the ability to resolve the smallest possible closely-spaced features and not just an isolated sharp edge at a particular section of the sample. For example, the smallest contacts and fins in integrated circuits[8], smallest inclusions in functional nanomaterials or the features of cellular sub structures[9] measure only a few-ten nanometer in size. Resolving such features is more challenging for table-top sources, compared to synchrotrons delivering similar flux, due to the longer wavelengths (>10 nm) typically employed. In consequence, the feature sizes are on the order of the wavelength and, for several important types of samples, effects like waveguiding play an important role in image formation[18].

In order to provide a reliable resolution characterization, the application of Rayleigh-type criteria on a standard Siemens star type of sample has been suggested for coherent microscopy[19]. A Siemens star provides small features in every in-plane direction which makes a comprehensive resolution analysis in all directions possible. In this work, resolving the smallest features in a table-top ptychographic imaging setup, with just a few wavelengths in size, is demonstrated for the first time. A Rayleigh-type criterion is used as a direct and reliable resolution metric in a transmission geometry that uses a Siemens star test sample.

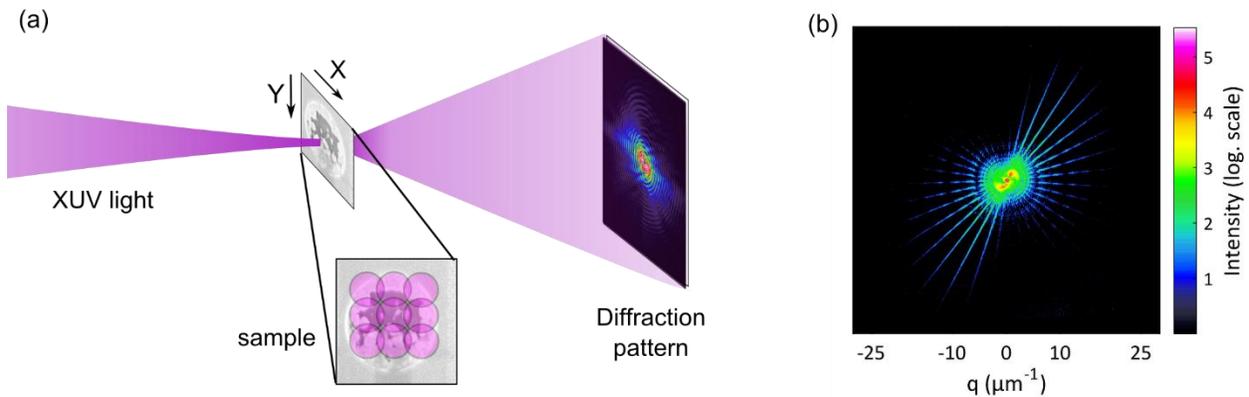

**Figure 1. Schematic diagram of a typical ptychography experiment. a**, an XUV or X-ray beam is focused on the sample to be imaged. The sample is raster scanned across the beam with sufficient overlap between consecutive illuminations and a diffraction pattern is recorded at each scan point. **b**, A representative high NA diffraction pattern recorded for the Siemens star sample used in this work (q is the momentum transfer).

An HHG source driven by a high average power fiber laser is used and a single harmonic line at 68.6 eV is selected and focused using spherical multilayer XUV mirrors (see Methods). An XY positioner with integrated position sensors is then used to translate the sample along the scan pattern and a diffraction pattern is recorded at each scan point (Fig. 1a). To limit the XUV beam size and have a constant illumination throughout the ptychographic scan, a pinhole with 6 µm diameter is placed ~400 µm upstream of the sample. A commercially available ultra-high resolution chart (NTT-AT:ATN/XRESO-20) is used as a sample and contains a Siemens star pattern with feature sizes down to 20 nm. Depending on the goal of the experiment, which can be high resolution or larger field of view, different values for the exposure time can be selected. For the high resolution measurement, two acquisitions with exposure times of 1 second and 10 seconds were taken at each scan point and stitched together for higher dynamic range (see Methods: data processing). One such composite diffraction pattern is shown in Fig. 1(b) and the Abbe limit from the geometry of the setup is 20 nm.

The ptychographic reconstruction is performed using the Difference Map[20] and the Maximum Likelihood algorithms[21]. Position correction algorithms[22] are also incorporated in the reconstruction to correct for sample/beam position drifts during the measurement (see Methods). The resulting reconstructed image is shown in Fig. 2(a). It shows multiple rings of the Siemens star pattern cleanly reconstructed over a field of view larger than 10 µm × 10 µm. The inset shows that the innermost ring of the Siemens star is resolved up to roughly half of its radius. A modified version of the Rayleigh criterion is used as a resolution metric to account for possible variations of resolution along different in-plane directions (see Methods). The resolution can be calculated by taking a circular cross-section along the innermost resolved lines (Fig. 2b). The half-pitch resolution (corresponding to the half-distance between two resolved features) is 45 nm in the vertical and 60 nm in the horizontal direction. In comparison, a knife-edge test on a feature with the sharpest 10% to 90% edge reveals 29 nm along the vertical and 45 nm along the horizontal direction. Obviously, the achieved resolution is not isotropic and the knife-edge criterion does not correctly resemble the true object space resolution measured by the Rayleigh criterion.

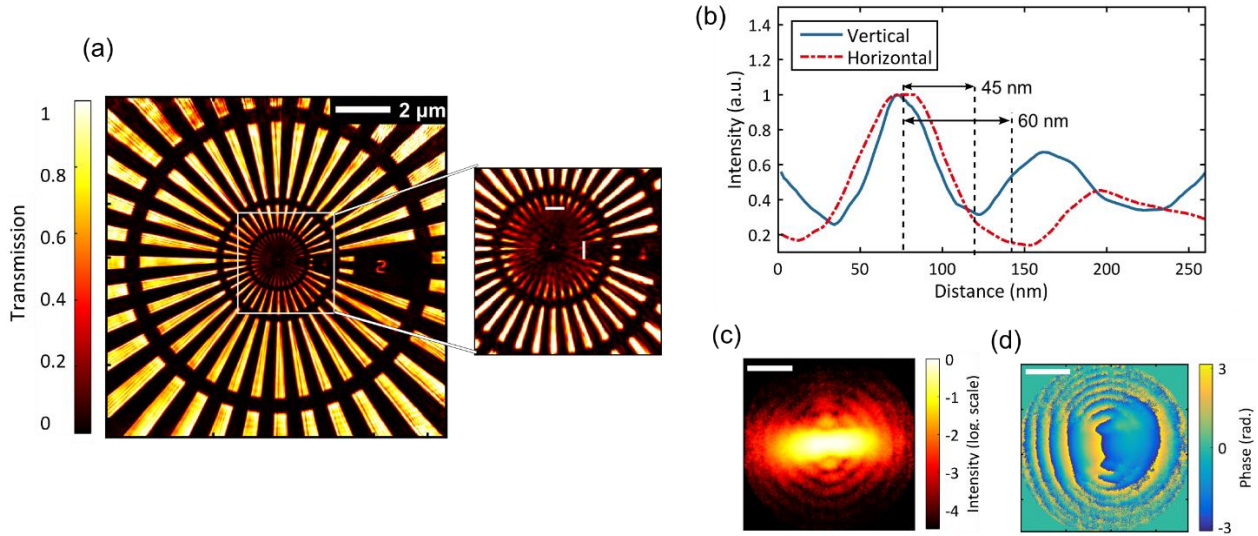

**Figure 2. High resolution ptychographic reconstruction. a**, reconstructed object over a field of view > 100 µm². The inset shows parts of the innermost ring (with smallest feature of 20 nm) being resolved. The number '1' in the inset with a size of 50 nm × 250 nm is resolved as well. **b**, cross-sections along the white vertical and horizontal bars in the inset of (a) demonstrate half-pitch resolutions of 45 nm and 60 nm. **c**, the reconstructed probe beam intensity shown in log. scale displays a horizontally elongated XUV beam. Multiple rings due to diffraction from the beam constraining pinhole are also visible. **d**, the reconstructed probe phase.

The anisotropy in resolution is caused by the dependence of the achievable resolution on the longitudinal coherence length of the source (~ 1.5 µm in our case) and the size of the probe beam[23]. The retrieved probe beam as shown in Fig. 2c has a width of 2 µm (1/e of the amplitude) in the vertical and 8 µm in the horizontal direction. The ellipticity of the probe beam is a result of the astigmatism introduced by the mirror arrangement used to focus the XUV beam on the sample (see Methods) and can easily be avoided by reducing the distance between sample and pinhole. This is merely a technical issue and can be solved in future by a more compact design of the XY-positioner and sample holder. In the current configuration, the ellipticity of the probe beam results in different temporal–coherence resolution limits along the horizontal (53 nm) and vertical direction (13 nm)[23]. Despite the astigmatism of the setup, the probe phase (Fig. 2d) is rather well defined for the high intensity region of the probe. Overestimation of the resolution from the knife-edge criterion compared to the Rayleigh-criterion suggests that numerical artifacts in the reconstruction can give rise to artificially sharp features not found in the actual sample. While this knife-edge method is well justified when used together with a Fourier-space analysis[24] of the highest spatial frequency consistently retrieved, it can lead to erroneous resolution estimates in the absence of one. These issues have been heavily discussed in the community[19] and a Siemens star test pattern with a Rayleigh-based criterion represents the best alternative for a fair comparison.

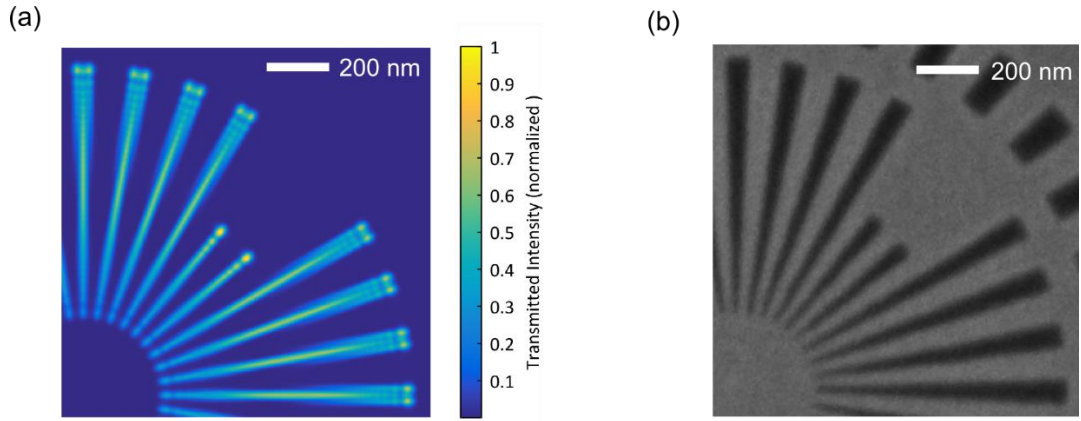

**Figure 3. Waveguiding effects limit the achievable resolution. a**, A finite difference time domain (FDTD) simulation of the innermost ring of the Siemens star pattern shows exit wave modulation and reduced transmission towards its center. **b**, Scanning electron microscopy (SEM) image of the same region of the sample as (a).

In addition, we identified another physical mechanism to limit the achievable resolution: waveguiding effects within the Siemens-star structure itself. These effects arise if the transverse sample structure size approaches the wavelength and the sample thickness is significantly larger. They can significantly modify the exit wave profile and reduce both the sample transmission and contrast[18,25]. A numerical simulation of wave propagation through the Siemens star pattern shows that the exit waves, behind the structured absorber, have a modulated and distorted intensity profile (Fig. 3a in comparison to the SEM image of Fig. 3b). These modulations couldn't be seen in the reconstruction due to their sub-pixel sizes. Moreover, the transmitted intensity and the contrast between spokes is reduced for decreasing structure size which hinders resolving of the smallest features. We additionally expect minor uncorrected fluctuations/drifts of the XUV beam which could limit the achievable resolution even though a beam stabilization system has been implemented. Obviously, resolving wavelength-scale features is significantly more difficult than detecting a sharp edge, which has been state-of-the art in table-top coherent imaging up to now[15–17].

The results presented up to now have been obatined with 20 minutes of total measurement time. Some applications might require a faster scan of a large area to identify regions of interest, which subsequently can be imaged with highest resolution. The measurement speed has been increased accordingly, by taking only a single diffraction pattern with reduced exposure time (3 sec.) at each scan point. The result of such a measurement with a scanning speed of 30 µm²/minute of a larger area (> 25 µm × 25 µm) is shown in

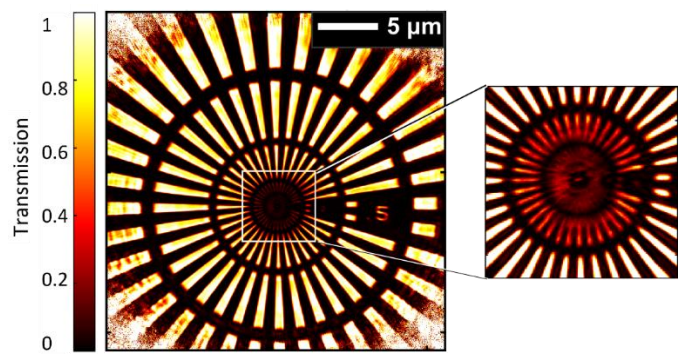

**Figure 4. Reconstruction with higher scan speed.** Reconstructed object over a field of view exceeding 600 µm² with a speed > 30 µm²/minute. The inset shows the innermost ring not resolved but still with non-zero transmitted intensity.

Fig. 4(a). The lines in the innermost ring (inset of Fig. 4(a)) are not resolved due to the lower acquisition time used in this case although their non-zero transmission compared to the opaque central region of the Siemens star is clearly visible. The smallest features of the second ring (with 200 nm pitch) are clearly resolved and the half-pitch resolution is thus at least 100 nm. Previous results on table-top setups didn't report exact scanning speeds but we estimated the speed of our measurement to be comparable to the fastest measurements[16]. Note that the sample's membrane has < 2% transmission at 68.6 eV since it is designed for hard X-rays. A more transparent membrane would significantly improve the resolution or reduce the required measurement times.

In summary, by imaging a Siemens-star test pattern we resolved the smallest features ever obtained by any table-top ptychography setup. A Rayleigh-resolution of 45 nm has been achieved with a field of view larger than 10 µm × 10 µm. In contrast to the typically employed knife-edge resolution test, the applied Rayleigh criterion is a direct object-space resolution metric and provides a reliable characterization of the imaging setup, which is an important prerequisite for future applications. In high speed mode, the scanning speed was increased to 30 µm$^2$/min for a larger field of view at a half-pitch resolution of ~ 100 nm. Since the photon energy used here is within the silicon transparency window, imaging of semiconductor structures, integrated circuits and mask inspection can directly benefit from this result. The compact light source and imaging setup will make nanoscale imaging with XUV light available to a larger scientific community. Moreover, the ultrashort XUV pulse durations of HHG sources will enable real-time observation of ultrafast processes. In the future, the achievable resolution will be improved by utilizing sources at shorter wavelengths which have already demonstrated photon flux of > 10$^9$ photons/s/eV for photon energies > 120 eV [10,26,27]. At the same time more compact laser sources will be available, which potentially could even be integrated on a machine-level e.g. for online wafer inspection or even in a clinical environment.

## Methods

### Experimental Setup

A high-order harmonic generation (HHG) source driven by a high-power femtosecond fiber laser system at 1030 nm with pulse energy of 0.6 mJ at 36 kHz repetition rate was used in the experiment. The HHG source delivered XUV radiation with energies up to 80 eV at a photon flux of up to 10$^{11}$ photons/second[28]. A single harmonic line at 68.6 eV was selected and focused at the sample area by a pair of off-axis multilayer spherical mirrors[29]. The relative bandwidth ($\Delta\lambda/\lambda$) of the illuminating XUV beam was estimated from previous measurements to be ~ 1/150 and was limited by the linewidth of the harmonic line. The off-axis focusing arrangement introduced astigmatism to the XUV beam. A small angle of incidence of 2° on the focusing mirrors was used to minimize this effect. The Full Width at Half Maximum (FWHM) XUV beam diameter was estimated to be ~ 10 µm at the focus (circle of least confusion). Therefore, a pinhole with 6 µm diameter was placed few hundred microns upstream of the sample to constrain the illuminating beam to the required size. The circle of least confusion of the astigmatic focus was placed at the beam-constraining pinhole to achieve maximum transmission. The XUV beam propagated about 400 µm after the circle of least confusion to reach the sample, resulting in an elongation of the beam along the horizontal direction. The pinholes were fabricated by focused ion beam (FIB) milling of a 200 nm gold layer on a 50 nm thick silicon nitride membrane. The commercial ultra-high resolution chart used in this work (NTT-AT:ATN/XRESO-20) had a 100 nm thick Ta absorber on top of a three-layer membrane made of 20 nm of Ru, 200 nm of SiC and 50 nm of Si$_3$N$_4$. The membrane had an overall transmission of 1.7% at 68.6 eV. The sample and the pinhole were mounted on independent XYZ piezo positioners (Smaract SLC-1740s) that are equipped with position sensors for closed-loop position control. An active beam stabilization system was incorporated in the HHG setup to stabilize the focus of the driving laser during the HHG process to < 4 µm position fluctuation and < 2 µrad angular fluctuation. This theoretically corresponds to < 100 nm position fluctuation and < 1 µrad angular fluctuation of the XUV beam focus. An XUV camera (Andor iKon-L, 2048 pixels and 13.5 µm/pixel) was placed ~ 27 mm after the sample to record the diffraction pattern.

## Data Analysis and Reconstruction

Different acquisition settings were used in the experiment depending on whether high resolution or larger field of view is desired. The ptychographic reconstruction was then performed using the Ptypy framework[30] where different algorithms (e.g. Difference Map (DM) and Maximum Likelihood (ML)) are implemented and additional algorithms can be easily incorporated. For the high resolution measurement shown in Fig. 2, the scan path consisted of 51 positions in a spiral pattern with a distance of 1.5 µm between two consecutive positions. At each scan point, two acquisitions with exposure times of 1 second and 10 second were taken. A 200 µm beam stop was employed for the 10 second measurement to block the central bright portion of the diffraction pattern and prevented the detector from saturating and blooming of the neighboring pixels. A dataset with higher dynamic range was constructed by merging the two measurements after scaling the 1 second dataset by an appropriate factor. The sample-camera distance was 27.6 mm resulting in an Abbe limit of ~ 20 nm. For the reconstruction, 3300 iterations of DM were used followed by 950 iterations of the ML algorithm. Position correction algorithms were applied in the last 600 iterations of the DM algorithm. For the measurement with a larger field of view shown in Fig. 4, a single acquisition with 3 seconds of exposure time was taken at each of the 200 scan points. The camera pixels were cropped during acquisition and the Abbe limit is 51 nm. In this case, 1500 iterations of the DM algorithm and 750 iterations of the ML algorithm were used for the reconstruction. The scanning speed is calculated as total scanned area divided by total measurement time. Total measurement time includes exposure time, readout time and the time required for moving the sample positioner from one scan point to the next.

## Resolution Estimation

Since the classical Rayleigh criterion refers to point like objects, but not to extended structures and since we had varying resolution values along different axes, we use a slightly modified resolution definition that identifies the minimum radius of the Siemens star where features are resolved. As suggested for standardized resolution test[19], the expected number of peaks of the Siemens star have to be identified at a certain radius to take a cross-section for resolution estimation (36 peaks within 360° angle for our sample). To account for the varying resolution along the horizontal (X) and vertical (Y) directions, we restricted the angles considered to ± 20° of +X, +Y, -X and –Y (four peaks within an angle of ± 20° in every direction). For both the horizontal and vertical directions, the minimum radius where the lines of the Siemens star are resolved were identified and cross-sections were taken. The larger value among the +X and –X cross-sections was taken as the achieved resolution along the horizontal and the same was done for the vertical direction. Two features are classified as resolved if the contrast

$$C = \frac{I_{peak} - I_{valley}}{I_{peak}}$$

between the peaks and the valley exceeded 0.264 [31]. In case of two peaks with differing intensity, the worst case scenario was considered by using the peak with lower intensity to calculate the contrast.

## Finite Difference Time Domain (FDTD) simulation

The FDTD simulation was performed using a commercial software package (Lumerical FDTD solutions). Only the innermost ring of the Siemens star was included in the simulation because of the high resource requirements of the rigorous method. A 100 nm Ta absorber layer on top of a 50 nm silicon nitride membrane was simulated in a three dimensional geometry with a uniform grid size of 0.75 nm. A plane wave with linear x-polarization was incident from the Ta side of the sample and the total field was monitored 9 nm behind the absorber inside the membrane. A total-field/scattered-field source was employed to realize an infinitely extended plane wave illumination within a finite computational domain. Outgoing scattered waves were absorbed by perfectly matched layer boundary conditions. The optical properties of Ta and silicon nitride were taken from the X-ray database of the Center for X-ray Optics[32].

**Acknowledgment**

The Authors acknowledge support from the Bundesministerium für Bildung und Forschung (BMBF) (05P2015), Federal State of Thuringia, European Social Fund (ESF) (2015 FGR 009) and Fraunhofer Gemeinschaft within Forschungscluster "Advanced Photon Sources". R.K. and V.S acknowledge support from the German Science Foundation DFG, IRTG 2101.

**References**

1. Sakdinawat, A. & Attwood, D. Nanoscale X-ray imaging. *Nat. Photonics* **4,** 840–848 (2010).

2. Miao, J., Ishikawa, T., Robinson, I. K. & Murnane, M. M. Beyond crystallography: diffractive imaging using coherent x-ray light sources. *Science* **348,** 530–5 (2015).

3. Rodenburg, J. M. *et al.* Hard-x-ray lensless imaging of extended objects. *Phys. Rev. Lett.* **98,** 34801 (2007).

4. Thibault, P. *et al.* High-Resolution Scanning X-ray Diffraction Microscopy. *Science (80-. ).* **321,** (2008).

5. Guizar-Sicairos, M. *et al.* High-throughput ptychography using Eiger: scanning X-ray nano-imaging of extended regions. *Opt. Express* **22,** 14859–70 (2014).

6. Shapiro, D. A. *et al.* Chemical composition mapping with nanometre resolution by soft X-ray microscopy. *Nat. Photonics* **8,** 765–769 (2014).

7. Holler, M. *et al.* X-ray ptychographic computed tomography at 16 nm isotropic 3D resolution. *Sci. Rep.* **4,** 3857 (2015).

8. Holler, M. *et al.* High-resolution non-destructive three-dimensional imaging of integrated circuits. *Nature* **543,** 402–406 (2017).

9. Jiang, H. *et al.* Quantitative 3D imaging of whole, unstained cells by using X-ray diffraction microscopy. *Proc. Natl. Acad. Sci. U. S. A.* **107,** 11234–9 (2010).

10. Rothhardt, J. *et al.* 53 W average power few-cycle fiber laser system generating soft x rays up to the water window. *Opt. Lett.* **39,** 5224–7 (2014).

11. Odstrcil, M. *et al.* Ptychographic imaging with a compact gas-discharge plasma extreme ultraviolet light source. *Opt. Lett.* **40,** 5574–7 (2015).

12. Klas, R. *et al.* Table-top milliwatt-class extreme ultraviolet high harmonic light source. *Optica* **3,** 1167 (2016).

13. Teichmann, S. M., Silva, F., Cousin, S. L., Hemmer, M. & Biegert, J. 0.5-keV Soft X-ray attosecond continua. *Nat. Commun.* **7,** 11493 (2016).

14. Popmintchev, T. *et al.* Bright coherent ultrahigh harmonics in the keV x-ray regime from mid-infrared femtosecond lasers. *Science* **336,** 1287–91 (2012).

15. Zhang, B., Gardner, D. & Seaberg, M. High contrast 3D imaging of surfaces near the wavelength limit using tabletop EUV ptychography. *Ultramicroscopy* (2015).

16. Baksh, P. D. *et al.* Wide-field broadband extreme ultraviolet transmission ptychography using a

high-harmonic source. *Opt. Lett.* **41,** 1317 (2016).

17. Gardner, D. F. *et al.* Subwavelength coherent imaging of periodic samples using a 13.5 nm tabletop high-harmonic light source. *Nat. Photonics* **11,** 259–263 (2017).

18. Zayko, S. *et al.* Coherent diffractive imaging beyond the projection approximation: waveguiding at extreme ultraviolet wavelengths. *Opt. Express* **23,** 19911 (2015).

19. Horstmeyer, R., Heintzmann, R., Popescu, G., Waller, L. & Yang, C. Standardizing the resolution claims for coherent microscopy. *Nat. Photonics* **10,** 68–71 (2016).

20. Elser, V. Phase retrieval by iterated projections. *J. Opt. Soc. Am. A* **20,** 40 (2003).

21. Thibault, P. & Guizar-Sicairos, M. Maximum-likelihood refinement for coherent diffractive imaging. *New J. Phys.* **14,** 63004 (2012).

22. Maiden, A. M., Humphry, M. J., Sarahan, M. C., Kraus, B. & Rodenburg, J. M. An annealing algorithm to correct positioning errors in ptychography. *Ultramicroscopy* **120,** 64–72 (2012).

23. Veen, F. van der & Pfeiffer, F. Coherent x-ray scattering. *J. Phys. Condens. Matter* **16,** 5003–5030 (2004).

24. van Heel, M. & Schatz, M. Fourier shell correlation threshold criteria. *J. Struct. Biol.* **151,** 250–262 (2005).

25. G.K. Tadesse, W. Eschen, R. Klas, V. Hilbert, D. Schelle, A. Nathanael, M. Zilk, M. Steinert, F. Schrempel, T. Pertsch, A. Tünnermann, J. L. and J. R. High resolution XUV Fourier transform holography on a table top. *Sci. Rep.* accepted (2018).

26. Hädrich, S. *et al.* Single-pass high harmonic generation at high repetition rate and photon flux. *J. Phys. B At. Mol. Opt. Phys.* **49,** 172002 (2016).

27. Ding, C. *et al.* High flux coherent super-continuum soft X-ray source driven by a single-stage, 10mJ, Ti:sapphire amplifier-pumped OPA. *Opt. Express* **22,** 6194–202 (2014).

28. Rothhardt, J. *et al.* High-repetition-rate and high-photon-flux 70 eV high-harmonic source for coincidence ion imaging of gas-phase molecules. *Opt. Express* **24,** 18133 (2016).

29. Tadesse, G. K. *et al.* High speed and high resolution table-top nanoscale imaging. *Opt. Lett.* **41,** 5170 (2016).

30. Enders, B. & Thibault, P. A computational framework for ptychographic reconstructions. *Proc. R. Soc. A Math. Phys. Eng. Sci.* **472,** 20160640 (2016).

31. E. H. K. Stelzer. Contrast, resolution, pixelation, dynamic range and signal-to-noise ratio: fundamental limits to resolution in fluorescence light microscopy. *J. Microsc.* **189,** 15–24 (1998).

32. CXRO X-Ray Interactions With Matter. Available at: http://henke.lbl.gov/optical_constants/. (Accessed: 20th January 2018)